\documentclass[twoside]{ilcws10}
\usepackage[latin1]{inputenc}
\usepackage[dvips]{graphicx,epsfig,color}
\usepackage{wrapfig,rotating}
\usepackage{amssymb,amsmath,array}

\begin{document}
\title{
%%%%   Paper title goes here  %%%%%%%%%%%%%%
Development of ultra-light pixelated ladders for an ILC vertex detector} %% 
%***********************************************************************
% AUTHORS INFORMATION AREA
%***********************************************************************
\author{N.~Chon-Sen$^{*1}$, J.~Baudot$^{1}$, G.~Claus$^{1}$, R.~De~Masi$^{1}$, M.~Deveaux$^{5}$, \\ 
W.~Dulinski$^{1}$, M.~Goffe$^{1}$, J.~Goldstein$^{4}$, I.-M.~Gregor$^{3}$, Ch.~Hu-Guo$^{1}$ \\
M.~Imhoff$^{1}$, C.~M\"untz$^{5}$, A.~Nomerotski$^{2}$, C.~Santos$^{1}$, \\
C.~Schrader$^{5}$, M.~Specht$^{1}$, J.~Stroth$^{5}$, M.~Winter$^{1}$.
% Optional short acknowledgment: remove next line if non-needed
%\thanks{This is an optional funding source acknowledgment.}
% DO NOT MODIFY THE FOLLOWING '\vspace' ARGUMENT
\vspace{.3cm}\\
% Addresses and institutions (remove "1- " in case of a single institution)
\vspace{.3cm}\\
1- IPHC/IN2P3/CNRS and Universit\'e de Strasbourg, Strasbourg, France \\
2- Sub-department of Particle Physics University of Oxford, Oxford OX1 3RH - UK \\
3- DESY Hamburg, Notkestr. 85, D-22607 Hamburg, Germany \\
4- University of Bristol, Tyndall Avenue, Bristol BS8 1TL, UK \\
5- IK-Frankfurt, Goethe-Universit\"at, Frankfurt am Main, Germany \\
}
%%***********************************************************************
% END OF AUTHORS INFORMATION AREA
%***********************************************************************

\maketitle

\begin{abstract} % 100 mots maximum
The development of ultra-light pixelated ladders is motivated by the requirements of the ILD vertex detector at ILC. 
This paper summarizes three projects related to system integration. 
The PLUME project tackles the issue of assembling double-sided ladders. %52
The SERWIETE project deals with a more innovative concept and
consists in making single-sided unsupported ladders embedded in an extra thin plastic enveloppe. %85
%These two projects both relies on the use of Monolithic Active Pixel Sensor (MAPS) 
%which integrates on the same silicon substrate the radiation sensor element (thin epitaxial layer) with processing 
%electronics (based on CMOS transistors) on top.
%Excellent minimum ionizing particle tracking performance has been demonstrated with 
%these devices on a series of MIMOSA (standing for Minimum Ionizing MOS Active sensor) 
%chip prototypes [1-3], with measured spatial resolution down to one 
%\mu m for 10 \mu m pixel pitch [4, 5].
AIDA, the last project, aims at building a framework reproducing the experimental running conditions
where sets of ladders could be tested. %106

\end{abstract}

%***********************************************************************************
\section{Introduction}
%***********************************************************************************
%\begin{wrapfigure}{r}{0.5\columnwidth}
\begin{figure}[h]
\vspace{-20pt}
\centerline
{
   \includegraphics[width=0.5\columnwidth]{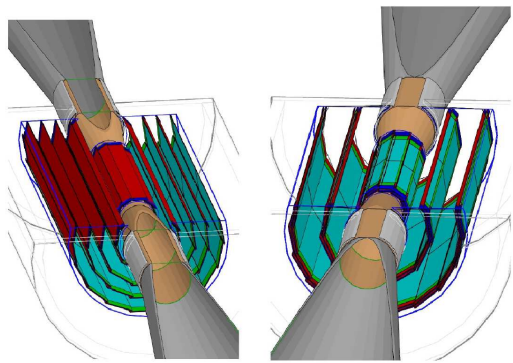}
}
\caption{The two geometry options for the ILD vertex detector. 
On the right a rather conservative approach based on five equidistant single-sided layers. 
On the left a more ambitious design with three layers equipped on both sides with sensors ($\simeq$ 2 mm apart).}\label{Fig:ILC_vertex_det}
%\end{wrapfigure}
\end{figure}
%A ladder consists of several butted sensors.
%The ILC vertex detector will be composed of several layers of ladders surrounding the interaction point.
%Figure \ref{Fig:ILC_vertex_det} shows the two envisaged geometry options for the International Linear Collider (ILC) vertex detector
%either composed of 5 layers of single-sided ladders (on the left) or 3 layers of double-sided ladders (on the right). \\
Physics goals and running conditions envisaged at the ILC
impose very strong requirements on the performances of the vertex detectors, whose two alternative geometries 
envisaged for the ILD setup \cite{LOI} are shown 
in Figure \ref{Fig:ILC_vertex_det}. 
In particular, the impact parameter resolution has to be as good as $\sigma_{IP} = a \oplus b/ (p\sin^{3/2}(\theta))$ 
where $a \leq$ 5 $\mu$m and $b \leq$ 10 $\mu$m GeV, far beyond what was achieved until now. 
This impacts strongly the allowed material budget, ambitiously defined in the ILD Letter of Intent as 
0.16\% and 0.11\% radiation length per layer for the double and the single-sided options respectively. \\
%We have to develop a concept to build such object which has to be mechanically stable and 
%to find partners from industries and laboratories to develop the expertise in building such objects. \\
%The compatibility with running conditions is also an important point. In the ILC environment, ladders will have to handle 
%an electromagnetic field $\simeq$ 4~T. 
%The sensor alignment will play an important role
%as it will determine to which extent we will benefit from the intrinsic good sensor resolution.
%The advantages of double-sided ladders versus the single-sided ones will be investigated
%as both options are being considered today. \\
Monolithic Active Pixel Sensors (MAPS) are developed at IPHC-Strasbourg to achieve this goal.
MAPS integrate on the same silicon substrate the radiation sensitive element (a thin epitaxial layer paved with p-n junctions) 
and the signal processing 
electronics (based on CMOS transistors).
Excellent minimum ionizing particle tracking performances were obtained with 
these devices on a series of MIMOSA (Minimum Ionizing MOS Active sensor) chips (\cite{MAPS1}-\cite{MAPS3}), 
with a measured spatial resolution down to 1 $\mu$m for a 10 $\mu$m pixel pitch (\cite{MAPS4},\cite{MAPS5}). 
They naturally offer a reduced material budget since the epitaxial layer
thickness amounts to less than 20 $\mu$m, resulting into a total useful sensor thickness below 30 $\mu$m. \\
Three projects related to sensor integration will be described in this paper: 
the PLUME project focuses on the concept of double-sided ladder,
the SERWIETE project addresses the issue of embedding sensors in an extra thin enveloppe to build a supportless single-sided ladder
and the AIDA project aims at building a framework where ladders 
can be tested in running conditions reproducing those of real experiments.

%***********************************************************************************
\section{PLUME project}
%***********************************************************************************

	\subsection{Introduction}
	\label{sec:plumeintro}

PLUME stands for \textit{Pixelated Ladder with Ultra-Low Material Embedding} (\cite{PLUME1},\cite{PLUME2}). 
The project has started in 2009 and aims at providing a double-sided ladder prototype 
with a material budget below 0.3\% radiation length in view of the ILD 
Detector Baseline Document to be delivered in 2012.\\ 
%It is the first integration project of MIMOSA sensors.
%\begin{wrapfigure}{r}{0.9\columnwidth}
\begin{figure}[h]
%\centerline{\includegraphics[width=0.30\columnwidth]{schema_principe_echelle_plume.eps}}
\includegraphics[width=0.9\columnwidth]{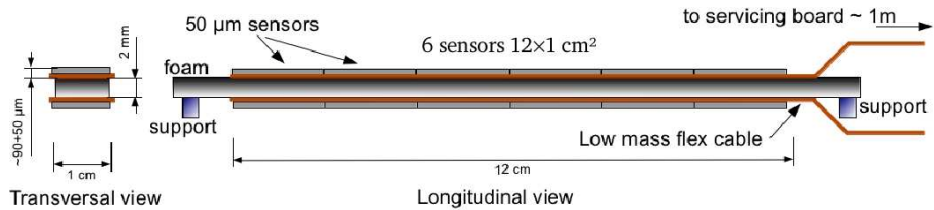}
\caption{PLUME ladder concept: 2 $\times$ 6 MIMOSA-26 sensors thinned down to
50 $\mu$m, plus low mass flex cables and foam support. The targeted material
budget is $\simeq$ 0.3\% radiation length.}\label{Fig:schema_principe_echelle_plume}
%\end{wrapfigure}
\end{figure}
%Figure \ref{Fig:schema_principe_echelle_plume} illustrates the PLUME ladder concept. 
In the double-sided option, two hits on the two sides of a ladder, 
originating from the same traversing particle, can be correlated and
used to reconstruct a \textit{minivector}.
Hence a better resolution on the impact parameter, a probably easier alignment and an improved 
reconstruction of shallow angle tracks.\\% ($<$15-30$^{\circ}$). \\ 
%There are several studies to be lead with PLUME in perspective of the DBD in 2012.
The PLUME collaboration encompasses the universities of Bristol and Oxford, 
DESY and IPHC (in synergy with IK-Frankfurt and LBNL).\\
The double-sided ladder concept, developed within this collaboration, is based on 
MIMOSA-26 sensors \cite{MIMOSA26} (18.4 $\mu$m pitch, 576$\times$1172 pixel array, 
Equivalent Noise Charge (ENC) $<$ 15 electrons, spatial resolution $\sigma$ $<$ 4 $\mu$m) thinned down to 50~$\mu$m.
%It is the first integration project of MIMOSA-26 sensors where 6 sensors will run simultaneously on the same flex cable.
Nevertheless the ladder concept developed is also adapted to other sensors such as Fine Pixel CCD (FPCCD) \cite{FPCCD} 
or In Situ Storage Image Sensors (ISIS) \cite{ISIS}.\\
Figure \ref{Fig:schema_principe_echelle_plume} illustrates the concept.
Six sensors are glued and butted on a kapton-metal flex cable.
To make a double-sided ladder, two flex cables are glued on each side of a stiffener, 
currently Silicon Carbide foam with 8\% density.
The flex cables are wire bonded to other, simpler, flex cables which transfer the signals and 
make the connection to the DAQ and servicing board located about 1~m away. \\
Sensors will be power-pulsed to benefit from the ILC beam cycle (at most 1/50 duty cycle) 
in order to reduce the average power consumption (around 100 mW/cm$^2$).
The whole setup will be air-cooled in order to ensure the optimized running conditions of the sensors. \\
%The total material budget for a PLUME ladder is estimated to be about 0.3\% radiation length. 
%Within this project several important points will be studied such as the impact of power pulsing 
%and air-flow cooling on the final sensor spatial resolution, the compliance with power cycling in a 3.5 T magnetic
%field and the benefits of double-sided ladder concept.
The main objectives of this work are the material budget suppression, that will be achieved in several steps, 
and the necessary expertise to produce the ladders (electrical design, mechanical support). 
The power consumption is also a crucial point that may be tackled by power pulsing the sensors (in a magnetic field)
and using air-flow cooling. The resulting indesirable consequences (Lorentz forces, vibrations) 
will be studied and suppressed in order to minimize their influence on the impact parameter resolution. \\
Another possibility for the double-sided ladder option would be to have sensors with different functionalities on each side.
One side could be dedicated to spatial resolution whereas the other would be optimized for temporal resolution.\\
The PLUME project is organised in 3 to 4 steps which allow to approach its target specifications in a progressive way.

	\subsection{First prototype: PLUME 2009}
	\label{sec:plume2009}
\begin{figure}[h]
\vspace{-20pt}
\begin{center}
%\centerline{\includegraphics[width=0.30\columnwidth]{schema_principe_echelle_plume.eps}}
\includegraphics[width=0.8\columnwidth]{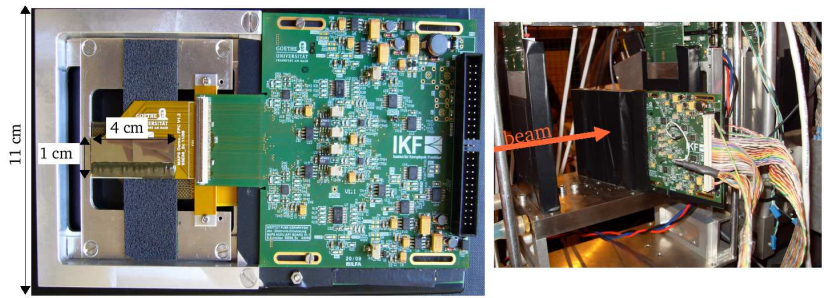}
\caption{Picture of the first PLUME ladder prototype (left) and its installation on the beam line in November 2009 (right).}\label{Fig:plume2009}
%\end{wrapfigure}
\end{center}
\end{figure}

\begin{wrapfigure}{r}{0.5\columnwidth}
\vspace{-40pt}
%\begin{figure}[h]
%\centerline{\includegraphics[width=0.30\columnwidth]{schema_principe_echelle_plume.eps}}
\includegraphics[width=0.48\columnwidth]{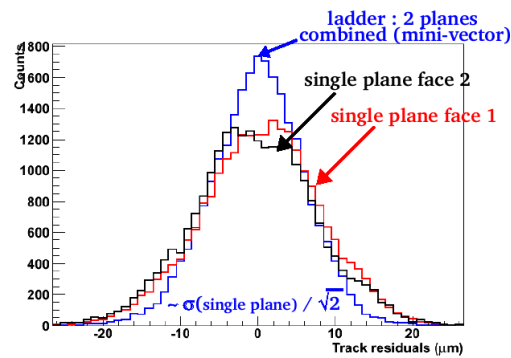}
\vspace{-10pt}
\caption{Track residuals derived from the beam data collected at CERN.}\label{Fig:track_residual_2009}
\vspace{-10pt}
\end{wrapfigure}
%\end{figure}
The goal of the PLUME 2009 ladder prototype was to settle the fabrication and beam test procedure.
It is therefore a still rudimentary version of the final device.
Figure \ref{Fig:plume2009} shows the prototype and the test setup that was tested in  
November 2009 with a 120 GeV pion beam at the CERN-SPS.
This prototype is composed of 2$\times$MIMOSA-20 analogue output sensors on each side, thinned down to 
50~$\mu$m, providing a 1$\times$4~cm$^2$ large sensitive area. 
A preliminary study was performed eventhough the sensors and the flex cables were not yet optimized.  
The results show as expected a 20 to 30\% improved spatial resolution (factor $1/\sqrt{2}$) by associating the hits on the two sides.
Figure \ref{Fig:track_residual_2009} displays in black and red the track residuals for each single plane ($\simeq$ 7.7 $\mu$m)  
and in blue those for the correlated hits ($\simeq$ 5.9 $\mu$m).

	\subsection{Next prototypes : PLUME 2010 and onwards}
	\label{sec:plume2010}
\begin{figure}[h]
\vspace{-20pt}
%\centerline{\includegraphics[width=0.30\columnwidth]{schema_principe_echelle_plume.eps}}
\includegraphics[width=1.\columnwidth]{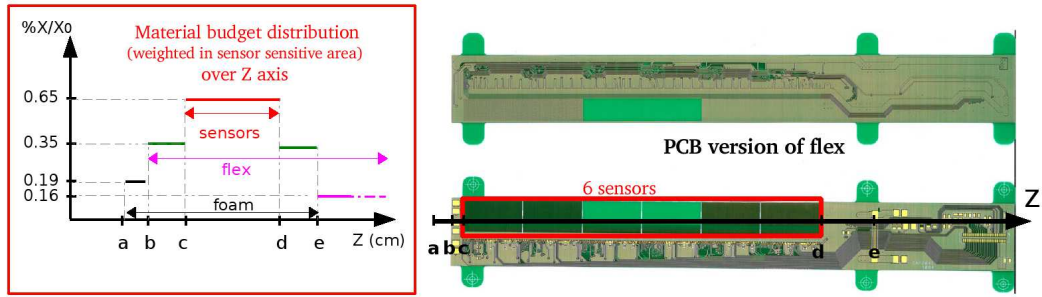}
\caption{PLUME 2010 material budget across the sensors sensitive area (left) and bottom and top view 
of a PCB version of the PLUME flex cable (right).}\label{Fig:plume2010_mat_bug}
%\end{wrapfigure}
\end{figure}
The 2010 prototype will be the first version featuring the final device design with 
6$\times$MIMOSA-26 binary output sensors operated simultaneously on the same flex cable.
The material budget estimate for this prototype is about 0.65\% radiation length in the sensors sensitive area.
Figure \ref{Fig:plume2010_mat_bug} displays the variation of material budget across the flex cable.
A ladder is expected to be mounted in Summer 2010 and tested on beam later on. \\
Several test benches are now being prepared. An air-flow cooling test bench at IPHC will enable 
to study the temperature spread over and through the flex cable in order to define the air-flow cooling system.
A power pulsing test bench is now operational at DESY and will study the different
 parameters governing the power consumption and allowing to reduce it without degrading the sensor performances.
A position survey test bench is also being prepared by Oxford and Bristol Universities.  \\
The 2011 flex design will tackle the issue of the material budget by reducing the traces thickness 
(currently 17 $\mu$m), adjusting the flex cable width to the sensors width (13.8 mm) 
and minimizing the impact of the stiffener.

%Presently everything is in time.
%Details on the project status are given in this list where I mention the different project deliverables and 
%the collaborators contributions.
%On the picture you can see a PCB version of the designed flex.

%REFAIRE CONLCUSION
%* plusieurs axes de développement pour PLUME pour 2012
%évidemment budget ma reduction mais du coup faut 
%develop expertise fabrication (design flex, support mécanique)
%develop acquisition pour les 6 capteurs => pb crosstalk, flot de données
%develop power pulsing => diminuer puissance dissipée
%develop systeme de refroidissement => vibrations
%develop etude champ magnétiques
%* Concept qu'on veut exporter à d'autres capteurs
%* AIDA

%The main objectives in view of the DBD in 2012
%are the material budget and integration time decrease.
%Within the years more advanced CMOS sensors will be used to achieve this goal.
%The perspectives for the PLUME project is to be studied within the infrastructure envisaged for the AIDA project.
%It may also have some interests for LHC experiment.
%This ladder concept will be developed in order to be able to accommodate other sensor types such as ISIS or FPCCD.

%***********************************************************************************
\section{SERWIETE project}
%% section headers !
%***********************************************************************************
%\begin{enumerate}
%\item \texttt{ilcws10\_template.tex} -- the text file
%\item \texttt{ilcws10.cls} -- the \LaTeX\ class file
%\item \texttt{ilcws10.eps} -- an example figure 
%\item \texttt{ilcws10\_template.pdf} -- the resulting pdf: author instructions
%\end{enumerate}
\begin{figure}[h]
\vspace{-20pt}
%\centerline{\includegraphics[width=0.30\columnwidth]{schema_principe_echelle_plume.eps}}
\includegraphics[width=1.\columnwidth]{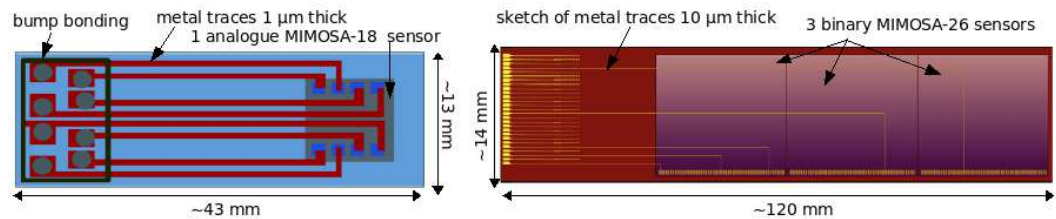}
\caption{Sketch of the two SERWIETE prototypes. 
On the left, the 2010 prototype equipped with 1 analogue output MIMOSA-18 sensor.
On the right, the 2011 prototype with 3 binary output MIMOSA-26 sensors with indication of some metal traces.}\label{Fig:serwiete_prototypes}
%\end{wrapfigure}
\end{figure}
The acronym SERWIETE stands for \textit{SEnsor Raw Wrapped In an Extra Thin Envelope}.
This project is part of the Hadron Physics 2 (HP2) E.U. program (FP7)
and aims at realizing an unsupported layer of thin MIMOSA sensors wrapped in polymerized 
film with a budget material below 0.15\% radiation length. \\ 
Because of their thin sensitive volume, 
MAPS can be thinned down to less than 30 $\mu$m, without affecting their tracking performance. 
This allows not only for a minute material budget but also for non-planar detector layers: 
thin silicon is flexible enough and can be bent to form a cylinder with a radius in the order 
of a centimeter. However this object is difficult to handle and to connect using standard 
methods (wire bonding). \\
A new challenging packaging technique is developed by the IMEC company (\cite{IMEC1},\cite{IMEC2}).
In this approach, the sensor is protected from both sides by polymer components
(10-20 $\mu$m thick, polyimide based insulator). Metal traces are deposited on top and directly 
connected to bonding pads through integrated vias in the polymer to provide electrical connection. \\
Two prototypes, as illustrated by Figure \ref{Fig:serwiete_prototypes}, are foreseen within this study.
The high-precision tracker MIMOSA-18 \cite{MIMOSA18} (10 $\mu$m pitch, 512$\times$512 pixel array, 
ENC $<$ 10 electrons, $\sigma <$  1 $\mu$m) has been chosen for the first monolithic sensor 
embedding process demonstrator. 
The first SERWIETE prototype will feature one metal layer and a silicon equivalent thickness of $\sim$50 $\mu$m.
It is expected to be fabricated by the end of 2010. 
The second prototype foreseen for 2011, will be equipped with 3 $\times$ MIMOSA-26 binary output sensors 
requiring two metal layers. 
Studies of the sensor tracking performances (charge collection properties, ENC, signal-to-noise ratio for minimum ionizing particles) 
as a function of mechanical stress (bending) will be performed as well as studies of the thermal properties and aging of the
packaging.
This embedding process may be extended to larger areas (tens of cm$^2$). 
It may also pave the way for the construction of cylindrical multi-sensors compact layers ($\simeq$ 10 mm thick) 
composing vertex detectors.

%Besides its minimal material content, the SERWIETE ladder presents the advantage of being flexible, 
%thus it can be installed on non planar surfaces acting as a mechanical support as for example the beam pipe. \\

%***********************************************************************************
\section{Alignment studies within the AIDA framework}
%% section headers !
%***********************************************************************************
\begin{figure}[h]
\vspace{-20pt}
%\centerline{\includegraphics[width=0.30\columnwidth]{schema_principe_echelle_plume.eps}}
\includegraphics[width=0.8\columnwidth]{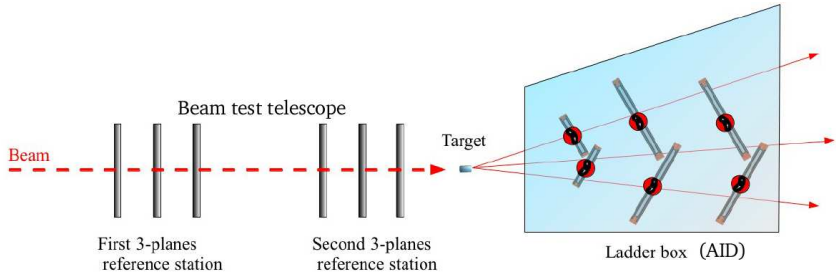}
\caption{Vertex oriented infrastructures for the AIDA project.}\label{Fig:aida}
%\end{wrapfigure}
\end{figure}

PLUME and SERWIETE address the problematics of realizing ultra-light sensor integration structures in order to minimize 
the material budget and thus the degradation of the micrometric spatial resolution provided by the sensors.
Another major issue for the vertex detector is the alignment.
To tackle it, infrastructures providing running conditions reproducing those of real experiments 
have been proposed within the European FP7 program AIDA \cite{AIDA}. 
The latter aims at delivering by 2013 infrastructures enabling to test sets of ladders. 
The infrastructure design takes into account the possibility of using different types of ladders.\\
The complete on-beam test infrastructure, as illustrated by Figure \ref{Fig:aida}, 
is composed of two Large Area beam Telescope stations (LAT) 
and of a ladder box called AID for \textit{Alignment Investigation Device}. 
The AID box contains three consecutive pairs of overlapping double-sided PLUME ladders
in an arrangement similar to the one of the ILD vertex detector. 
The two double-sided ladders belonging to a pair can translate and rotate with respect to each other 
in a controlled way with a micrometric precision.
A removable target will be placed in front of the box 
in order to study the vertex reconstruction capabilities. \\
An off-beam test infrastructure will investigate thermo-mechanical aspects of the detector design
and their influence on the alignment including the effect of an air-flow cooling system and the impact
of power cycling in a strong magnetic field.

%\begin{wraptable}{l}{0.5\columnwidth}
%\centerline{\begin{tabular}{|l|r|}
%\hline
%Plenary talk  & 1 page per 3 minutes talk \\\hline
%Parallel talk  & 1 page per 3 minutes talk  \\
%\hline
%\end{tabular}}
%\caption{Page count guidelines.}
%\label{tab:limits}
%\end{wraptable}
%Figure \ref{Fig:MV} shows an example of a figure and related

%***********************************************************************************
\section{Summary}
%***********************************************************************************

System integration studies of low mass pixelated vertex detectors are under way 
for the ILD Detector Baseline Document to be delivered in 2012. 
To achieve the ambitioned impact parameter resolution, the material budget of the ladders
composing the detector should not exceed a few per-mill of radiation length. Within the ILD apparatus, the ladders may be equipped with
sensors on their two sides in order to improve the overall vertexing performances. \\
The development of ultra-light ladders is ongoing within the PLUME and SERWIETE projects.
The questions addressed are prominent generic issues susceptible to deteriorate the sensors 
micrometric resolution, such as the ladder material budget, the compliance with
a strong magnetic field and an air-flow cooling system. 
These developments will also allow investigating the added value of double-sided ladders as well as the alignment issue.
The latter will be addressed within the infrastructure of the recently approved European project AIDA.

%***********************************************************************************
\section{Bibliography}
%***********************************************************************************
%If possible please use the bibtex information as given by SPIRES
%to make the citations~\cite{parton_qed} uniform and follow the 
%examples~\cite{parton_qed,H1,DVCS,pomeron} given below.
%Note that there is a (non-breaking) space before \verb?\cite?.

% ****************************************************************************
% BIBLIOGRAPHY AREA
% ****************************************************************************

\begin{footnotesize}
% IF YOU DO NOT USE BIBTEX, USE THE FOLLOWING SAMPLE SCHEME FOR THE REFERENCES
% ----------------------------------------------------------------------------

% ----------------------------------------------------------------------------

\end{footnotesize}

% ****************************************************************************
% END OF BIBLIOGRAPHY AREA
% ****************************************************************************

\end{document}